\documentclass[a4paper,11pt]{article}
\begin{document}
\def\boxit#1{\vcenter{\hrule\hbox{\vrule\kern8pt
      \vbox{\kern8pt#1\kern8pt}\kern8pt\vrule}\hrule}}
\def\Boxed#1{\boxit{\hbox{$\displaystyle{#1}$}}} 
\def\sqr#1#2{{\vcenter{\vbox{\hrule height.#2pt
        \hbox{\vrule width.#2pt height#1pt \kern#1pt
          \vrule width.#2pt}
        \hrule height.#2pt}}}}
\def\square{\mathchoice\sqr34\sqr34\sqr{2.1}3\sqr{1.5}3}
\def\Square{\mathchoice\sqr67\sqr67\sqr{5.1}3\sqr{1.5}3}
%
\title{\bf Ether and Relativity}
\author{Mehrdad Farhoudi\thanks{E-mail:
 m-farhoudi@sbu.ac.ir}\ \ and Maysam Yousefian\thanks{E-mail: M\_Yousefian@sbu.ac.ir}\\
 {\small Department of Physics, Shahid Beheshti University,
         Evin, Tehran 19839, Iran}}
\date{\small November 29, 2015}
\maketitle
\begin{abstract}
\noindent
 We consider one of the fundamental debates in performing
the relativity theory, namely, the ether and the relativity points
of view, in a way to aid the learning of the subjects. In
addition, we present our views and prospects while describing the
issues that being accessible to many physicists and allowing
broader views. Also, we very briefly review the two almost recent
observations of the Webb redshift and the ultra high--energy
cosmic rays, and the modified relativity models that have been
presented to justify them, wherein we express that these
justifications have~not been performed via a single model with a
single mechanism.
\end{abstract}
\medskip
{\small \noindent
 PACS number: $03.30.+p$\ ; $95.85.Ry$\ ; $04.20.Cv$\ ; $98.80.Jk$}\newline
 {\small Keywords: Ether Theory; Relativity Theory; Absolute Space;
                   Lorentz Violation; Varying Speed of
                   Light; Doubly Special Relativity.}
\bigskip
\medskip
\bigskip

In commemorating the first century of the discovery of general
relativity by Albert Einstein that was recognized as a triumph of
the human intellect, it would be instructive to look through one
of its fundamental debates, namely, between the {\it ether} and
the {\it relativity} points of view. Certainly, very vast amount
of work have been performed on these subjects and the references
given in this compact survey are~not obviously a complete
bibliography on these topics, and although we provide adequate
references, it is a self--contained work. However, while we are
trying to spell out some basic issues behind the subject, the work
almost provides a brief review in a different perspective about
the long history and the situations of the ether and relativity up
to the present days. Nevertheless, it has~not been only aimed to
give just a motivation to research on the issue, and we propose,
somehow during the work, to introduce our points of view and
prospects on these subjects in a way that being accessible to many
physicists and allowing broader views.

It seems that it was Descartes who first introduced into science
the concept of ether as a space--filling material in the manner of
a container and a transmitter between distant bodies (similar to
what that, nowadays, we call it a field) in the first half of
seventeenth century~\cite{ref27-2}. About one generation after
him, perhaps one can consider Newton as one of the ether theory
pioneers who practically introduced ether into physics. Actually,
Newton presented the concept of {\it inertia} in the first law
(i.e., the inertia law) of his famous three laws of
mechanics~\cite{Newton}, and in this respect, he considered an
inertial frame as a rigid frame in which free particles move with
constant speed in straight lines. On the other hand, a free
particle is a particle that moves with constant speed in a
straight line in an inertial frame; and obviously, this is a {\it
vicious circle} (or, a logical loop). In another word, it is
ambiguous that what thing distinguishes or singles out the class
of inertial frames as criteria or standards of non--acceleration
from all other frames. Newton, who was also aware of this
difficulty, in order to specify the inertial frames from
theoretical point of view, employed the idea of {\it absolute}
space with the aid of the notion of ether. He considered absolute
space as a {\it rest} inertial frame (or the Newtonian ether) that
actually is a very thin motionless media with nearly zero density,
perfect luminosity and strong elasticity character, which is also
a conveyer for force transmission. To Newton's contemporaries,
like Hooke and Huyghens, the ether's main function was just to
carry light waves and thus, it could also be acted
on~\cite{ref27-2}. However, the Newton idea on the ether based on
it as an acting substance which does~not accept reaction, but
Leibniz (or, Leibnitz) insisted that the space is an order of {\it
coexistences}\footnote{This point of view is a return to the
Aristotle idea. Incidentally, this idea is related to the {\it
relational physics}~\cite{Michael-2008,Dean-2008}, which means a
physical system is in a way that positions and other properties of
things have meanings just with respect to the other things. This
point of view is a prelude to the Mach ideas, particularly the
{\it weak version} of it.}\ \cite{Erlichson-1967,Capek-1976}. He
argued against the Newton idea of {\it substantival ontology of
space}, and believed that this idea leads to contract with the
{\it principle of sufficient reason}~\cite{Earman1987}. Also,
Berkley presented some arguments against the Newton absolute space
on his work named {\it De Motu} (On Motion)~\cite{Berkley-1721}.
However, when the luminiferous ether evolved into a cornerstone of
the Maxwell theory~\cite{ref000}, it became a plausible marker for
the Newton absolute space.

In Newtonian physics, space is a pre--existing stage on which
material particles are the characters acting out the drama of
physical events. This point of view is on the contrary to the
Aristotelian\footnote{The Aristotle point of view on space was
asserted in his definition of {\it place} as {\it the limit
between the surrounding and the surrounded
body}~\cite{Adler-1966}, and also as {\it the innermost motionless
boundary of that which surrounds
it}~\cite{Hardie-Aristotle,Aristotle}.}\
 view that space is a {\it plenum} (i.e.,
occupied by matter) and {\it inseparably} associated with the
material substance~\cite{Hardie-Aristotle,Aristotle}. In fact, the
Newton view is a return to the Democritus view that space is a
{\it void} with the properties which are {\it independent} of the
material bodies that move {\it in} it~\cite{Adler-1966}. While in
relativistic gravitational physics, again, space cannot be
considered apart from the matter that is in it, and, as the
mathematician E.T. Whittaker~\cite{Adler-1966} points out, in this
case, the characters create the stage as they walk about on it,
i.e., gravitation has become part of the stage instead of being a
player. In another word, the properties of space in gravitational
theories are inseparable from the matter that is in it. Indeed, it
has been pointed out~\cite{Mashhoon-1994} that a basic problem of
Newtonian mechanics is that the {\it extrinsic} state of a point
particle, i.e. its appearance in space and time (that usually
characterized by its position and velocity), is a \emph{priori}
independent of its {\it intrinsic} state (that usually
characterized by its mass). However in quantum physics, each
coordinate (or in another word, position that is the notion of
geometry) does~not commute with its corresponding momentum (or in
another word, dynamics that can be considered as the notion of,
moving, mass); or in other words, for each object, these two
characters are~not simultaneously compatible from an observer
view. That is, analogous to the {\it complementarity} principle of
the particle--wave duality, the issue may be interpreted as in
confrontation with everything, it either represents the aspect of
geometry or the aspect of matter in one instant depending on the
experimental arrangements and/or the initial conditions.

Nevertheless, and principally, the innovation of absolute space is
while the Galilean transformations do~not distinguish among the
inertial frames as well, and thus Newton, in confronting the
quarry that how absolute space can be specified, presented the
famous idea of the Newton {\it bucket} from the practical point of
view~\cite{d'Inverno-1992,Janssen2005}. However, the Newton bucket
provides the distinction of a non--inertial frame, and does~not
distinguish the inertial frames from each other. That is, any
curve or change in the horizontal level of the bucket water does
represent the acceleration of the bucket with respect to a frame
which is itself either an inertial or a non--inertial one.
However, Newton accounted it with respect to an inertial frame (as
a criterion to distinguish acceleration), and actually with
respect to a specified inertial one, i.e., absolute space.
Although in this regard, Mach also interpreted the changes with
respect to the average motions of all particles in the universe
(or, the distant fixed stars)~\cite{Mach1977}. Indeed, and in a
more accurate expression, the Newton discussion was that any curve
of the horizontal level of the water cannot be because of its
relative rotation with respect to the bucket, however, Mach did
consider it as the relative motion between
them~\cite{Barbour-1995}. Incidentally, and up to the available
experiments, one may also~not being able to locally detect an
accelerated frame in the large scales, e.g., the rotation of the
earth around the sun and or the rotation of the solar system
around the center of our local galaxy, merely by the idea of the
Newton bucket.

Before we continue our discussions, it would be also instructive
to review the following well--known proposed experiment on the
issue. Consider the rotation of the plane of a swinging Foucault
pendulum at the earth's north pole. Within the limits of
experimental accuracy\rlap,\footnote{Also, the {\it
Lense--Thirring precession} effect (see, e.g.,
Refs.~\cite{LenseThirring}--\cite{Ciufolini2000}), or actually the
{\it frame--dragging} effect~\cite{GP-B,Iorio2011}, must be
neglected.}\
 the remarkable fact is that the times taken for the earth to
rotate a complete round with respect to absolute space, and
relative to the fixed stars are the same. In Newtonian view, there
is nothing a {\it priori} to predict this result, and it is just a
{\it coincidence}. In other words, the result indicates that the
fixed stars are~not rotating (or, do~not have acceleration)
relative to absolute space, and can be employed as a criterion to
specify the class of inertial frames. However in Machian view, one
precisely expects that the two time durations of the measurements
must be the same regardless of the accuracy of the instruments,
for, in his view, the detected criteria of acceleration are
exactly the fixed stars.

In the historical process, Maxwell offered an another way to
specify the Newton ether. In the ether theory presented by
him~\cite{ref000}, absolute space is accounted as a media for the
light propagation, and is specified via the Maxwell equations (or
from his point of view, the ether equations). Regarding this
theory, due to the {\it Fresnel dragging}
effect~\cite{French-1966,Resnick-1968}, the speed of light in the
other inertial frames is different from the ether one, as the
Maxwell equations are~not invariant under the Galilean
transformations. To investigate the Maxwell ether, the {\it
Michelson--Morley experiment}~\cite{ref00-1}-\cite{Shankland2} was
performed in the year of 1887, and the Maxwell ether was~not
confirmed. However, to explain the null result of the
Michelson--Morley experiment, Fitzgerald proposed a hypothesis in
the year of 1889~\cite{ref00-1-1}. According to his hypothesis,
when a body moves with a constant velocity with respect to the
ether, it will be (really) contracted in the direction of motion.

Lorentz, like Maxwell, believed that the light propagation,
similar to the sound, requires a media which is the ether as the
characteristic of absolute rest. Thus, he also
proposed~\cite{ref00-2,ref00-3} a hypothesis (although
independent, but actually an elaboration of the Fitzgerald
hypothesis) to explain the Michelson--Morley experiment. According
to the Lorentz ether theory~\cite{Lorentz1909}, a body with a
constant velocity with respect to the ether is also
contracted\footnote{Such a contraction was accounted in terms of
the Lorentz electron theory~\cite{Lorentz1909}; however, it is
believed that some other results predicted from his theory
could~not be found experimentally~\cite{Resnick-1968} and the
theory has some philosophical deficit such that its basic
assumptions are unverifiable~\cite{d'Inverno-1992}.}\
 in the direction of motion (the
{\it Lorentz--Fitzgerald contraction}) and its clocks are slowed
down~\cite{Lorentz1904} (i.e., the clock retardation and/or (real)
time dilation) when moving through the ether. By his theory, the
results of the Michelson--Morley and even the {\it
Kennedy--Thorndike}~\cite{ref27-1} experiments, which is a broader
and more general one than the previous experiment, are also
explained. Incidentally, based on the Lorentz ether theory, and
similar to the Newtonian mechanics, the inertial reference frames
are related to each other by the Galilean transformations, however
according to the Lorentz--Fitzgerald contraction hypothesis, the
form of the Maxwell equations still remain invariant. In this
case, as the earth is surrounded by the ether, there are two
choices. Either the ether must be dragged by the earth and remains
at rest with respect to it, which the {\it aberration
observation}~\cite{ref0-1,Stewart-1964} has rejected this choice.
Or, the ether must~not be dragged by the earth and contains a
velocity with respect to it, in a way that the Fresnel dragging
should be observed for light. This subject was investigated via
the {\it Fizeau experiment}~\cite{ref0-3}--\cite{ref0-4}, and the
Lorentz ether theory has also been able to explain the result of
this experiment by the aid of local time dilation hypothesis.
Regarding these facts, there have been some debates, comparisons
and investigations about the originality and the equivalence of
the Lorentz ether theory with the special theory of relativity
({\bf STR}); see, e.g.,
Refs.~\cite{ref27-2,Ives}--\cite{Erlichson} and references
therein. However, the two theories are logically independent,
because obviously, the choice of different postulates principally
leads to theories which differ in their simplicity and appeal,
although they may observationally be equivalent.

However, Einstein, and Mach before him, were also among the
opponents of the Newton absolute space, and in this regard, they
raised two main objections. First, how absolute space, as an
inertial frame, can be theoretically distinguished and located
from all other inertial frames in a unique way. Second, how
absolute space can act on every particle and distinguishes free
particles from other ones, but cannot be acted upon. In general,
from the Einstein point of view, the existence of a matter which
is completely transparent, its nature is unspecified and obscure,
and there is no way to prove its existence\rlap,\footnote{However,
nowadays, the discovery of the acceleration of the
universe~\cite{Riess-1998}-\cite{Riess2004} can be discussed as a
possible way of investigating the contrary to such a claim.}\
 was~not required. Eventually, in 1905, Einstein by proposing
the STR~\cite{Einstein-1905}, attracted most of the attentions
towards this theory. Actually, Einstein presented his theory by
generalizing the Newton absolute space to the Minkowski
spacetime~\cite{Minkowski} and extending Newtonian relativity with
the Galilean transformations to special relativity with the
Lorentz transformations, and could explain the whole mentioned
experiments very well. However in the STR, the inertial frames, as
the preferred ones, are still the references and the criteria for
the absoluteness of the concept of
acceleration\rlap,\footnote{Thus, the Newton first law is
consistent with special relativity. However, for distinguishing
the inertial frames from the rigid frames, instead of the existing
context of the Newton first law, one can employ, e.g., the law of
light propagation. In fact, the usual definition of rigid bodies
cannot be applied in special relativity, although to be consistent
with it, some new definitions have been stated. For instance, the
characteristic of rigidity is assigned to a body as
relative--rigidity that {\it any length element of the body on the
move remains invariant with respect to the comoving
observer}~\cite{Born} or {\it a body on the move somehow deforms
continuously that each of its infinitesimal elements has just the
Lorentz contraction with respect to the instantaneous rest
observer}~\cite{Ehrenfest}.}\
 and yet, the difficulty remains in
theoretically distinguishing these class of frames from all other
frames. Indeed, the Galilean relativity principle, in general,
contains the four--dimensional special relativity
formulation~\cite{Petkov2005}. Nevertheless, in the STR, as the
equivalence of the inertial frames are valid for all physical laws
(including the Maxwell equations), the Maxwell ether hypothesis is
rejected. However, it cannot verify absolute space (although,
still has the mentioned objections to it), but it cannot also deny
its existence even though absolute space cannot be distinguished
by intrinsic properties from all other inertial frames. On this
point of view, Einstein was~not satisfied with the theory too.

It may be surprising, but, perhaps due to some points (like the
mentioned ones), even Mach, whose critiques to the Newtonian
mechanics paved the way to the relativity theory smooth (at least
philosophically), was suspicious on the Einstein
theory\rlap.\footnote{His critical view on the STR has been
explicitly expressed in the foreward to the ninth edition of his
book~\cite{Mach1977}.}\
 However, due to many experimental verifications of the STR
obtained from the wide diversity of different phenomena, the
skeptics had to give up; indeed, special relativity probably is
the most based and reliable tested theory in the contemporary
physics. Nevertheless and despite the enormous experimental
robustness on the STR, in the last two decades, due to the
theoretical posited questions, scientists are eagerly looking for
experimental findings that somehow violate the
STR~\cite{Iorio2006}. In this regard, the researches are
particularly focused on experiments that indicate violation of the
Lorentz symmetry\rlap.\footnote{Among the reviews on the Lorentz
violation, Ref.~\cite{Mattingly-2005} includes the theoretical
approaches as well as the phenomenological analysis.}\
 However, and as a rough estimation, the quantum gravity induced
Lorentz violation can only be achieved as a theoretical purpose,
for the natural scale that one would expect (in this respect and
as a strong violation) is the Planck energy of about $10^{19}$
GeV, while the highest known energy particles is the ultra
high--energy cosmic rays ({\bf UHECR})~\cite{ref5}--\cite{ref6} of
about $10^{11}$ GeV and the present accelerator energies are about
$10^{3}$ GeV that preclude any direct observation of the Planck
scale Lorentz violation. Nevertheless and in particular, some
physicists have looked for evidence on invalidity of the principle
of relativity~\cite{Drozdzynski2011}, or for deriving special
relativity from Galilean mechanics alone~\cite{Sela2009} and or
for testing the STR by investigating what sort of new bounds can
be achieved at high energies while the Lorentz symmetry is~not
satisfied~\cite{ColemanGlashow}.

On the other hand, on the way to reject absolute space and
absolute concept of acceleration, Einstein himself also attempted
to propose the general theory of relativity ({\bf GTR}) in
1915~\cite{Janssen2005,Einstein-1915,Einstein-1916} inspired from
the Mach ideas~\cite{ref27-2,Barbour-1995,Lichtenegger-2005} and
with the aid of the {\it principle of equivalence} of gravitation
and
inertia~\cite{d'Inverno-1992,CiufoliniWheeler1995,Straumann2004}
as a clue principle\rlap.\footnote{Meanwhile, for grasping more
about the contents and points that led to the advent of the GTR;
see Refs.~\cite{Mehra}--\cite{Norton}.}\
 Of course, in the GTR, no way or solution has also been provided for
determining the inertial frames, and in fact in this theory, no
preference or reference, as a criterion of non--acceleration, is
taken to these frames as well. That is, it is appeared that in the
GTR, the question of how to determine the inertial frames has been
deleted. Nonetheless, in the GTR, contrary to the {\it strong
version} of the Mach principle (i.e., space is~not expressed as an
independent essence/substance, but merely as an abstraction from
the totality of distance--relations between matter), the spacetime
is attained as an independent essence/substance which both acts on
the matter and is reacted upon (indeed, it has the weak version of
the Mach principle). Meanwhile, the confusion surrounding the
principle of equivalence led a physicist like Synge to suggest, in
the preface of his book about the GTR~\cite{Synge1960}, that this
principle has to be set aside and the facts of absolute spacetime
be faced. Although, to clear some of the ambiguities about the
principle of equivalence, it is emphasized that its statement must
be stated {\it locally}, wherein by locally it is meant a region
over which the variation of the gravitational field cannot be
detected~\cite{d'Inverno-1992}. Nonetheless, even with this type
of statement, it seems that still the ambiguities have~not been
completely eliminated\rlap.\footnote{Principally in gravitational
physics, energy itself acts as a source of gravity and is~not
capable of simply being thrown away, and also one cannot easily
rescale the zero point of it~\cite{bida,bos}.}

However, about the two main raised objections on absolute space,
in the GTR for elimination of the first objection, there is no
need or preference requirement to propose an absolute space.
Resolution of the second objection is also expressed by accepting
an {\it independent essence} (or, substance) for geometry (i.e.,
space), and actually, by appealing to the weak version of the Mach
principle. On the other hand, while forming the GTR, it was
specified~\cite{Stachel-1912} via the {\it hole argument} (or,
problem)\footnote{It had been thought that generally invariant
field equations cannot uniquely determine the gravitational field
generated by certain distributions of source masses, in
contradiction with the requirement of physical
causality~\cite{d'Inverno-1992,Janssen2005,Einstein-1913,Einstein-1914}.}\
 that the point events of the spacetime manifold had been incorrectly
thought of as individuated independently of the field itself. That
is, it is impossible to drag the metric field away from a physical
point in empty spacetime and leave that physical point behind. As
Einstein himself wrote~\cite{Einstein-Ehrenfest,Einstein-Besso}
that nothing is physically real but the totality of spacetime
point coincidences, and placed~\cite{Einstein-1952} great stress
on the inseparability of the metric and the manifold. Hence, the
spacetime continuum (i.e., physical events) is the same as space
points (or, manifold) that are~not separated from the metric
(i.e., geometry of space)\rlap.\footnote{This is known as the {\it
point--coincidence} discussion.}\
 In fact, a
key lesson raised from the Einstein gravitational theory is that
the gravitational field has been inseparably twisted/intricated
with the geometry of spacetime, and thus, geometry itself is an
impellent essence (or, dynamic).

In this regard, it is worth to mention that most of the leading
relativists in the early twentieth century, for examples,
Eddington~\cite{Eddington-1921} and even Einstein
himself~\cite{ref27-4}, claimed that, in principal, the GTR is
merely an ether theory\rlap.\footnote{In this respect; see also
Ref.~\cite{Janssen2005} and references therein.}\
 On this issue, Trautman has asserted~\cite{Trautman-1966} that he
has presented the mathematical demonstration of such a claim by
obtaining a form of the GTR without spatial curvature. And
recently, by employing a combination of Lorentz's and Kelvin's
conception of the ether\rlap,\footnote{That is, the ether as a
substance of some kind, and not a type of vacuum without any
properties intrinsic to itself (e.g., the ether would have the
property of ponderability, which is to say, it has the power to
gravitate or to generate curvature).}\
 and actually by using the Lorentz--Kelvin ether
theory~\cite{Whittaker-1954,Whittaker-1951,Schaffner-1972}, the
Einstein field equations has been
obtained~\cite{Dupre-2012}\rlap.\footnote{By adopting that the
ether gravitates {\it only} in the presence of matter.}\
 Meanwhile, it has also been claimed in Ref.~\cite{Gautreau-2000} that
there is an underlying relationship between the GTR and Newton's
absolute time and space (via the existence of a preferred set of
coordinates in general relativity\footnote{Note that, in general,
any relativistic gravitational equation, including the Einstein
equations of the GTR, is needed to be non--linear, and hence, its
number of independent solutions are~not finite and the {\it
superposition principle} is~not also valid for it.}\
 that is equivalent
to Newton's absolute time and space). And even it has been
asserted~\cite{Savickas} that, in terms of the Newton laws within
4--dimensional curved geometries, the GTR can be exactly
described.

Also in relevance to the Mach ideas, some physicists just believe
that, in his ideas, the inertial frames have been replaced by the
average motions of all particles in the universe and the influence
of distribution of matter in the immediate vicinity of any
particle, as well as the other distant bodies, determines the
inertial frames. Nonetheless, in this kind of belief, this new
reference adapts its nature from the whole matter of the universe.
On the other hand, in a few recent decades, some novel ideas and
theories have been proposed as ``geometrical description of
physical forces'', ``geometrical base of material content of the
universe'', ``geometrical curvature induces matter'' and
``induced--matter theory'' which usually connect extra dimensions
(i.e., geometry) to the properties of the
matter~\cite{Salam-1980}--\cite{Rasouli-2014}. Even some
gravitational theories have been considered in which the
Lagrangian of the geometry, that is usually supposed to be the
characteristic of the geometry alone, just from the beginning, and
indeed {\it a priori}, is presented as a function of the geometry
and the matter~\cite{Harko-2008}--\cite{ZareFarhoudi}. Now,
inspiring from these types of ideas and theories, and knowing
that, in the Mach ideas, the inertia of a body is~not just the
intrinsic property of that body, but is {\it caused} by the cosmic
masses via some interactions (where the influence of the distant
bodies preponderates)\rlap,\footnote{However in the Mach ideas,
there is also no description about why the interaction should be
velocity--independent, but acceleration--dependent, and or indeed,
why there is such a distinction between unaccelerated and
accelerated motion in the nature.}\
 one can perhaps relate this {\it cause} to be
due to the cosmic background (e.g., the whole geometry of the
universe).

Nevertheless, to make the GTR more consistent with even the strong
version of the Mach principle, Einstein
inserted\footnote{Incidentally, by this insertion, he also
provided the possibility of having a {\it static} solution for the
universe (that was thought to be so at that time), as an
appropriate condition on the GTR.}~\cite{EinsteinCC} the
well--known term of the cosmological constant\footnote{Even a
sufficiently small value of the cosmological constant can have
very important effects on the evolution of the universe; and
although the implications of this term are cosmological, the
origin of it is probably to be found in the quantum theory rather
than cosmology.}\
 into his equations\rlap.\footnote{The Einstein equations are
$G_{\mu\nu}=(8\pi G/c^{4})T_{\mu\nu}$, where $G_{\mu\nu}$ is the
Einstein tensor as a function of the metric and its derivatives,
$G$ is the Newtonian gravitational constant, $T_{\mu\nu}$ is the
energy--momentum tensor and the lower case Greek indices run from
zero to three. The Einstein equations plus the cosmological
constant are $G_{\mu\nu}-\Lambda g_{\mu\nu}=(8\pi
G/c^{4})T_{\mu\nu}$, where $g_{\mu\nu}$ is the metric tensor and
$\Lambda$ is a constant.}\
 However, when de Sitter achieved~\cite{deSitter} his solution for
the vacuum GTR plus the cosmological constant
term\rlap,\footnote{Meanwhile and almost around the same time,
{\it non--static} closed solutions of the GTR (corresponding to an
expanding distribution of matter) were discovered, and also it was
specified that the universe is~not static, but rather is expanding
in the large--scale (that was officially published a few years
later~\cite{Hubble1929}).}\
 Einstein vehemently retook the inclusion of such a term while
describing it as the biggest mistake he ever
made~\cite{Gamov1970}. Even in this regard, realizing that the
metric field is~not a phenomenon resulted of matter but has its
own independent existence, Einstein, near the end of his life,
gradually decreased his enthusiasm for the Mach principles. Indeed
in 1954, he wrote to Pirani that one should no longer speak of the
Mach principles at all~\cite{Pais,Pirani}. Perhaps the main point
of the issue roots in considering spacetime as a new inertial
standard which directly influences by the active gravitational
mass through the Einstein equations, although, in the absence of
mass and other disturbances, still spacetime would straighten
itself out into the class of extended inertial frames, contrary to
the idea that all inertia is caused by the cosmic masses.

Nonetheless, by considering the necessity of {\it conformal
symmetry breaking}, the inclusion of the cosmological constant
term is still proposed to remedy the inconsistency of the Einstein
gravitational theory with the strong version of the Mach
principle~\cite{NamFar}. On the other hand, besides confronting
the cosmic gravitational collapse (due to gravity among them),
there needs to have a kind of repulsing force for explaining the
recent discovered acceleration of the
universe~\cite{Riess-1998}-\cite{Riess2004}, which seems to
originate from ``property" of geometry itself or spacetime in
global scales, contrary to the well--known forces up to now. In
this respect, the Einstein equations including the cosmological
constant term have again been considered, and this term is
interpreted as if the vacuum fluid and the vacuum energy
density\rlap,\footnote{Incidentally, according to quantum theory,
the vacuum has {\it vacuum fluctuations} and an energy tensor
(zero--point energy) that the only form of it (being the same in
all inertial frames) is a constant multiple of the metric, i.e.
the same as the cosmological constant term. However, the
calculations based on theories of elementary particles yield a
value for the corresponding cosmological constant term to be
orders of magnitude far larger than the observations allow. This
discrepancy is known as the {\it cosmological constant problem};
see, e.g., Refs.~\cite{Cos.pro1}-\cite{Bernard} and references
therein.}\
 see, e.g., Refs.~\cite{WeinbergBook,Barrow2011}. Incidentally in
this regard, the ether energy--momentum tensor introduced in
Ref.~\cite{Dupre-2012} is~not dissimilar to this term. Also in the
last two decades, in the {\it dark energy}
issue~\cite{Peebles-2003}--\cite{Bamba-2012} (an energy that
consists nearly $69\%$ of the matter density of the
universe~\cite{Ade-2013,Ade-2015}), it seems as if the geometry
(or in another word, space) in the cosmological scale has an
anti--gravity type of interaction. In essence, in these ideas,
both the geometry and matter (in its general meaning, including
material and radiation) are {\it different aspects} of a ``thing''
(or in another word, existence), although, even by accepting an
independent entity for each one, they would somehow relate to the
other one as well (at least, through that ``thing''); see, e.g.,
Ref.~\cite{Farhoudi-2006} and references therein.

In addition to the dark energy issue, the other cosmological
observations have indicated~\cite{dmatt1}--\cite{dmatt5} that
there should also be another kind of matter besides the usual
barionic matters, i.e. an exotic fluid called {\it dark matter},
that consists~\cite{Ade-2013,Ade-2015} nearly $26\%$ of the matter
density of the universe. These two important cosmological problems
and, on the other hand, the quantizing
difficulty~\cite{bida,bos,Farhoudi-2006,farc} of the Einstein
gravity (in spite of the impressive successes of it) are, in
general, the main reasons that have raised the need to investigate
generalized or alternative gravitational theories. In this
respect, and for instance, one of the alternative theories is the
Brans--Dicke gravitational theory~\cite{Brans-1961} that is more
consistent with the Mach ideas. In connection to our discussion,
also in this theory, there is a kind of matter in the form of a
scalar field in the whole space in addition to the usual matter
(or, the barionic matter)~\cite{Dicke-1962}. Actually, while the
Brans--Dicke gravity is regarded as the generalized Einstein
gravity, its Lagrangian can be converted to the Einstein
gravitational Lagrangian plus a scalar field term via the {\it
conformal transformation}~\cite{Fujii-2004,Farajollahi-2010}.
Meanwhile, in the other gravitational theories of type of the
Brans--Dicke gravity (or in general, the scalar--tensor
gravitational
theories~\cite{Fujii-2004,Faraoni-2004,Capozziello-2011}), in
particular, the {\it chameleonic} gravitational
theories~\cite{Khoury-2004}--\cite{SabaFarhoudi}, by the coupling
of a scalar field with the metric (or in another word, space), the
dynamic of the scalar field depends on the surrounding background
density which requires that the interaction of this scalar field
with the usual matter to be of gravitational type. Among different
types of the modified gravitational theories, one can also mention
the
Einstein--ether~\cite{Mattingly-2005,JacMat2001}--\cite{Haghani2014}
(and references therein) and non--minimal
{\ae}ther-–modified~\cite{Furtado-2013} gravity theories. In these
theories, in general, the coupling of the Einstein gravity with a
dynamical timelike vector field (representing a preferred rest
frame, i.e., ether) is considered.

Essentially, one of the three probable assumptions that Brans and
Dicke stated in their work~\cite{Brans-1961} is that physical
space has intrinsic geometry and inertial properties beyond those
that can be achieved from the matter contained therein, however,
in their work~\cite{Brans-1961}, they proceeded the other
assumption that leads to the Brans--Dicke gravitational theory.
Nonetheless, and also according to the Dicke
view~\cite{Dicke-1962}, the introduced scalar field in this theory
is a field that along with the metric are described as the
gravitational field (or in another word, geometry). In this
regard, in the ancient time, although Plato did~not accept the
view of {\it void} space and believed that space is a {\it plenum}
(i.e., a general assembly), but his view was also different from
the Aristotelian one. In the Plato view, space is an entity that
bodies are made out of it and cannot exist without
it~\cite{Adler-1966}\rlap.\footnote{This point of view seems not
to be irrelevant with the strong version of the Mach principle.}\
 In other words, Plato identified space as that in which things come to
be~\cite{Archer-Plato,Plato}.

Moreover, in the last two decades, another two observations,
namely the Webb redshift~\cite{Webb-1999,ref1} and the
UHECR~\cite{ref5}--\cite{ref6,Albert}, have been reported while
the standard Einstein relativity theory is~not capable to explain
these two cosmological phenomena. Hence, it was required that some
modifications and generalizations being performed on the Einstein
relativity. In this respect and up to now, there have been
represented several models to describe the Webb redshift,
including the models for varying the constants that participate in
determining the atomic structure~\cite{Barrow-1998}. Among these
types of models, one can mention the varying electric
charge~\cite{ref2} and the varying speed of light ({\bf
VSL})~\cite{Barrow-Magueijo-1998}--\cite{ref4-5} (and references
therein) models, where the comparison of these two kinds of models
has also been performed in Ref.~\cite{ref3}. On the other side,
along with theories such as the non--commutative field
theory~\cite{ref18,ref19}, the most reliable models, that attempt
to explain the observed UHECR, are known as the
doubly--special--relativity or deformed--special--relativity ({\bf
DSR})~\cite{ref10}--\cite{ref17}. However, all the available
modified models on these subjects have been unsuccessful in
justifying these two phenomena via {\it a single} model with a
single mechanism.

To clarify the latter expression, let us very briefly review how
these modified models work. Actually, it would be instructive to
represent a concise description of these two recent phenomena and
an overall explanations on the VSL and the DSR regarding
justifications of these two observations.

During the observations of galaxies and distant stars covering the
redshift range $0.5<z<3.5$, the Australian group of Webb noticed
redshifts that can be justified with a variable fine structure
constant~\cite{Webb-1999,ref1}. Actually, in the standard
cosmology, the ratio of the cosmological redshift (due to the
expansion of the universe) of the absorption lines spectra of
atoms on distant galaxies to the ones of the same atoms in
laboratory is predicted to be the same for different amounts of
energies of the absorption bands. However, Webb {\it et al.}
observed that this ratio depends on the quantum numbers and the
atomic and the molecular structures of the materials that radiate
the corresponding rays, and hence, the structure of the absorption
bands should vary due to the redshift caused by the expansion of
the universe. As in the standard cosmology, the redshift usually
means distant past, thus, the explanation of such a phenomenon has
been based on having different atomic structure (and hence, the
absorption bands) in distant past with respect to its present
structure. In this regard, among the models that aim to explain
the Webb redshift, the VSL models are the best option.

In relativity area and in general, the VSL models can usually be
classified into two methods. In one method, new scalar fields are
added to the Einstein--Hilbert Lagrangian, and another method is
mainly based on changing this Lagrangian itself. In general, the
appearance of any scalar field can be performed somehow to make
variation in the speed of light, for, naively, it is analogous to
have light rays passing through a dielectric media. It means that
the appearance of any dielectric media causes variation in the
speed of light, and if there is no dielectric in the matter media,
the constancy of the speed of light will be in the vacuum. Thus,
in these models, the speed of light practically depends on the
appearance of the scalar field, by which also, the other
cosmological issues, such as inflation, flatness and dark energy,
are usually described.

As a simple prototype, although general, a scalar--tensor action
for the VSL models, analogous to the one used in Ref.~\cite{ref2},
can be written as
\begin{equation}\label{eq0}
S=\int d^{4}x\sqrt{-g} \left(
L^{[g]}+L^{[\psi]}+L^{[m]}e^{-2\psi}\right),
\end{equation}
where $L^{[g]}\equiv R/16\pi G$ is the Einstein--Hilbert
Lagrangian,
$L^{[\psi]}\equiv-\omega(\psi)\partial^{\alpha}\psi\,\partial_{\alpha}\psi/2+V(\psi)$,
$V(\psi)$ is a self--interacting potential and $L^{[m]}$ is the
matter Lagrangian. Also, $R$ is the Ricci scalar, $\omega(\psi)$
is a varying dimensionless coupling coefficient of the scalar
field $\psi$, $g$ is the determinant of the metric and, for
simplicity, we have set the speed of light, in the absence of the
scalar field, to be $c\left(\psi=0\right)=1$. Variations of this
action, with respect to the metric and the scalar field, yield
\begin{equation}\label{eq0-a}
\Square\,
\psi=\frac{1}{\omega}\left(2e^{-2\psi}L^{[m]}-\frac{\omega
'}{2}\partial^{\alpha}\psi\partial_{\alpha}\psi-V'\right)
\end{equation}
and
\begin{equation}\label{eq0-b}
G_{\mu\nu}=8\pi G\left(
T^{[\psi]}_{\mu\nu}+T^{[m]}_{\mu\nu}e^{-2\psi}\right),
\end{equation}
where the prime denotes the ordinary derivative with respect to
the argument, $\Square\equiv {}_{;\,\rho}{}^{\rho}$ and
$T^{[i]}_{\mu\nu}\equiv -(2/\sqrt{-g})\delta (\sqrt{-g}\,
L^{[i]})/\delta g^{\mu\nu}$. Now, by employing the spatially flat
homogeneous and isotropic metric of the
Friedmann--Lema\^{i}tre--Robertson--Walker ({\bf FLRW})
\begin{equation}\label{eq-metric-FLRW}
ds^{2}=dt^{2}-a^{2}(t)\left( dr^{2}+r^{2}d\Omega^{2}\right)
\end{equation}
that includes the scale factor $a(t)$, the Friedmann--like
equation for a perfect fluid achieves as
\begin{equation}\label{eq0-c}
\left( \frac{\dot{a}}{a}\right) ^{2}=\frac{8\pi G}{3}\left(
\rho^{[m]}e^{-2\psi}+\rho^{[\psi]}\right),
\end{equation}
where $\rho^{[m]}$ is the matter density and with the assumption
of the scalar field being also homogeneous, we have
$\rho^{[\psi]}=\omega\dot{\psi}^{2}/2+V$. At last, using the
cosmological considerations, the resulted equations specify the
way that the scalar field and, in turn, the speed of light vary.

Meanwhile, in the VSL models, one should note that if the metric
is assumed to be
\begin{equation}\label{eq-metric-FLRW-mofat}
ds^{2}=c^{2}(t)dt^{2}-a^{2}(t)\left(
dr^{2}+r^{2}d\Omega^{2}\right),
\end{equation}
it cannot by itself being used as the variation of the local speed
of light (i.e., as the one that travels along the null geodesics),
and hence, as the local violation of the Lorentz symmetry.
Because, the speed of light, in the absence of matter, is just a
criterion of the variation of time with respect to the place,
which this kind of variation does~not have an interesting meaning.
Indeed, the variation of the speed of light and the local
violation of the Lorentz symmetry, on the Riemannian manifold, can
be considered only in two cases. Either the assumption is that
light rays travel on a manifold and observers are on another one,
wherein this case, light rays obviously do~not travel along the
null geodesics of the observers. As an example of this case, we
can mention the ``induced--matter''
models~\cite{Wesson-1999,Wesson-2006} and some of the
multi--metric models~\cite{Alexander-2000}. Or, as an another
case, there exist some fields on the Riemannian manifold that, by
interaction with light, prevent light rays traveling along the
null geodesics~\cite{ref23}--\cite{ref24}.

Among the VSL models, it is worths to mention the bimetric model,
e.g. Refs.~\cite{ref23}--\cite{ref24}, in which the effective
metric of light and matter, $\breve{g}_{\alpha\beta}$, is distinct
from the spacetime metric $g_{\alpha\beta}$ as
\begin{equation}\label{eq0-d}
\breve{g}_{\alpha\beta}\equiv
g_{\alpha\beta}+B\partial_{\alpha}\psi\,\partial_{\beta}\psi ,
\end{equation}
where $B$ is a constant coefficient with the dimension of the
inverse of the energy density. The corresponding action is
\begin{equation}\label{eq0-e}
S=\int d^{4}x\sqrt{-g} \left(
L^{[g]}+L^{[\psi]}+\frac{\sqrt{-\breve{g}}}{\sqrt{-g}}\breve{L}^{[m]}\right),
\end{equation}
where all the terms are as in action~(\ref{eq0}) except that here,
$\omega$ is constant and the matter Lagrangian is a function of
the effective metric $\breve{g}_{\alpha\beta}$. In this model,
light rays do~not travel along the spacetime geodesics, and thus,
there exist local variations of the speed of light with respect to
the speed of graviton~\cite{ref23}--\cite{ref24}. Then, variations
of the action, with respect to the metric and the scalar field,
yield
\begin{equation}\label{eq0-g}
\Square\,
\psi=\frac{1}{\omega}\left(B\frac{\sqrt{-\breve{g}}}{\sqrt{-g}}\breve{T}^{[m]\,\mu\nu}\,
\breve{\nabla}_{\mu}\breve{\nabla}_{\nu}\psi-V'\right)
\end{equation}
and
\begin{equation}\label{eq0-h}
G_{\mu\nu}=8\pi G\left(
T^{[\psi]}_{\mu\nu}+\frac{\sqrt{-\breve{g}}}{\sqrt{-g}}\breve{T}_{\mu\nu}^{[m]}\right).
\end{equation}
And again, for all the same mentioned conditions, the
corresponding Friedmann--like equation for a perfect fluid is
\begin{equation}\label{eq0-i}
\left( \frac{\dot{a}}{a}\right)^{2}=\frac{8\pi G}{3}\left(
\frac{\sqrt{-\breve{g}}}{\sqrt{-g}}\breve{\rho}^{[m]}+\rho^{[\psi]}\right),
\end{equation}
that eventually, with the aid of the cosmological considerations,
specifies how the scalar field and, in turn, the speed of light
vary by the time.

As mentioned, the other important recent observation, that is
considered as a challenge for the STR and the Lorentz symmetry, is
the UHECR~\cite{ref5}--\cite{ref6,Albert}. At first, let us
briefly describe this phenomenon. When particles usually reach the
specified energies, then, due to interaction with the microwaves
background (that can be the cosmic infrared
background~\cite{ref7-1,ref7} and or the cosmic microwaves
background~\cite{ref7-2}), either would be significantly absorbed
by the pair--production (like, for the high--energy photons), or
their energies are reduced via the photon--pion production (like,
for the ultra high--energy protons and neutrons). Hence, the ultra
high--energy cosmic particles have limited life--times, and thus,
can travel limited distances. In this regard, in the year 1966, in
two distinct papers~\cite{ref8,ref9}, the threshold energies were
specified for the distances that can be traveled by the ultra
high--energy cosmic particles (depending on the amount of their
energies) using the calculations based on the quantum field theory
and according to the Lorentz symmetry. These threshold energies,
that are derived via the STR, $E_{\rm th-SR}$, are known as the
{\bf GZK} threshold after Greisen--Zatsepin--Kuzmin. According to
these calculations, the threshold energy for the high--energy
photons are about $10^{4}$ GeV, and for the ultra high--energy
protons are about $5\times 10^{10}$ GeV~\cite{ref5}. Nevertheless,
the ultra high--energy protons and photons have been observed that
their energies are more than the corresponding calculated
threshold energies~\cite{ref5}--\cite{ref6}. On the other hand, as
there is no source for such ultra high--energy particles inside
our galaxy, hence, these particles have, in principle, been able
to travel extragalactic distances. Therefore, this observation
infers more life--times for these particles than the calculated
ones based on the Lorentz symmetry and the standard quantum field
theory~\cite{ref5-1,ref6}.

To explain the observation of the UHECR, as mentioned, the most
reliable approach is the DSR~\cite{ref10}--\cite{ref17}, wherein,
inspired from the notion of quantum gravity, it contains two
invariant scales, e.g., the speed of light and the Planck energy.
Of course, in the realm of DSR, different models have been
presented. Some of these models are based on the quantum
groups\footnote{For this subject; see, e.g.,
Ref.~\cite{Majid2000}.}\
 and the non--commutative geometry~\cite{ref11,ref16}, wherein, the
corresponding quantum groups are related to Hopf algebra. In these
type of models, either the corresponding relation of the
non--commutative geometry is one of the obtained results of the
quantum groups relations~\cite{ref11}, or based on the
non--commutative geometry assumption, the corresponding relations
of the quantum groups are resulted~\cite{ref16}. Some of the other
models of the DSR are presented based on the projective linear
group~\cite{ref12,ref13,ref14}. As a prototype of these models,
the model of Magueijo--Smolin~\cite{ref12} can be mentioned, in
which by substituting the Fock--Lorentz~\cite{Fock1964} inertial
transformations instead of the Lorentz ones, new transformations,
named Magueijo--Smolin, have been defined for the energy--momentum
space. Then, by the new transformations, a specified scale of
energy (for instance, the Planck energy) is set as an invariant
for different inertial observations. Also, some other models of
the DSR are stated based on the deformation in the generators of
the Lorentz group~\cite{ref10,ref17}.

However, in general, the common key point, in all of the
explaining models~\cite{ref18}--\cite{ref17} of the observation of
the UHECR, is that the linear equations of the field and also the
energy--momentum are somehow replaced by non--linear ones. For
instance, in Ref.~\cite{ref10}, the dispersion relation (that
usually specifies the connections between the energy, momentum and
mass through the Klein--Gordon equation, and also is the indicator
of the linear equation of wave) is modified by a length parameter
(for instance, the Planck length, $\ell_{\rm p}$) as a Lorentz
violation parameter. Thus, in this way, via the modified
dispersion relation and the conservation laws of energy and
momentum, the value of the threshold energy, $E_{\rm th-SR}$,
increases to a new value, $E_{\rm th}$. For example, in
Refs.~\cite{ref10,ref17}, the change in the dispersion relation
for an ultra--relativistic (i.e., $E\gg m$) particle of mass $m$
and energy $E$ has been considered, via the Lorentz violation
parameter and in leading order in the Planck length, as the
non--linear form
\begin{equation}\label{eq0-1}
E^{2}\simeq p^{2}+m^{2}+\varepsilon\,\ell_{\rm p}\, p^{2}E,
\end{equation}
where $p$ is the momentum of the particle and $\varepsilon=\pm 1$
depending on the model under consideration. Note that, as we have
set $c=1=\hbar$, the Planck length has the dimension of the
inverse of energy, and actually, the Planck energy is the
invariant (observer--independent) maximum energy scale similar to
the speed of light for the speed of particles. Also, although
relation~(\ref{eq0-1}) is~not invariant under the Lorentz
transformations, but it is under a sort of amended Lorentz ones
depended on the considered model of the DSR. Then, as the
threshold energy of a high--energy particle is the value of the
energy that the particle can interact with the microwaves
background, with the aid of relation~(\ref{eq0-1}) and the
conservation laws of energy and momentum before and after the
collision, the value of the threshold energy in the case of the
Lorentz symmetry, $E_{\rm th-SR}$, is amended up to the
first--order in the Planck length to be~\cite{ref17}
\begin{equation}\label{eq0-2}
E_{\rm th}+\varepsilon\,\ell_{\rm p}\frac{E_{\rm th}^{3}}{8E_{\rm
IR}}\simeq E_{\rm th-SR},
\end{equation}
where $E_{\rm IR}$ is the background infrared energy (soft--photon
energy) and $E_{\rm th}$ is the physical (amended) threshold
energy. Nonetheless, we should also mention that there are some
problems in the models of the DSR, such as the lack of a standard
approach for achieving the DSR and not having a unique type of
transformations of the spacetime, that have been considered in
Ref.~\cite{ref17}. Moreover, it has been
argued~\cite{Hossenfelder} that the DSR with an energy--dependent
speed of light has some inconsistencies, and wherein, the
present--day observations in particle physics rule out its
first--order modification in the speed of light.

Now, let us consider our above expression about these two types of
models. First, in the VSL models, those scalar (or vector) fields
(that are introduced to describe the variation of the speed of
light) essentially need to be assumed almost constants (or very
slowly varying) over small cosmological intervals, for to comply
with the recent cosmological considerations. Thus, these models
are in no use for justification of the observed UHECR which are
actually effective in those intervals. On the other hand, although
the DSR has emerged as a VSL effective model, but it acts just for
the observational implications of the
UHECR~\cite{Albert,Magueijo,ref23,ref41}. Indeed, with the speed
of light as a function of energy, it predicts the variation of the
speed of light in the range of ultra high--energies. Thus, such a
variation of the speed of light in the DSR is in no use for
justification of the observed Webb redshift which indicates the
variations of the speed of light in low--energies. Therefore,
there is no single model for justification of these two observed
phenomena via a single mechanism. In this regard, in another
work~\cite{YouFar}, we have defined and introduced a new type of
ether model, consistent with the Mach ideas, that can justify both
of the observed phenomena.

\section*{Acknowledgements}
\indent

We thank the Research Office of Shahid Beheshti University for the
financial support.

%

\begin{thebibliography}{1}
\bibitem{ref27-2}W. Rindler, ``\textit{Relativity: Special, General and
         Cosmological}", (Oxford University Press, Oxford, 2nd Ed. 2006).
\bibitem{Newton}I. Newton, ``{\it Philosophi{\ae} Naturalis Principia Mathematica}",
         (Streater, London, 1st Ed. 1687), Final Ed. in English by: A. Motte, 1729, Revised
         by: A. Cajori, ``{\it Sir Isaac Newton's Mathematical Principles of Natural Philosophy
         and His System of the World}", (University of California Press, Berkeley, 1962).
\bibitem{Michael-2008}F. Michael, ``\textit{Leibniz's Metaphysics of
         Time and Space}", (Springer, Heidelberg, 2008).
\bibitem{Dean-2008}R. Dean, ``\textit{Symmetry, Structure and Spacetime}", (Elsevier,
         Oxford, 2008).
\bibitem{Erlichson-1967}E. Erlichson, ``The Leibniz--Clarke controversy: Absolute versus
         relative space and time", \textit{Am. J. Phys.} \textbf{35} (1967), 89.
\bibitem{Capek-1976}M. \v{C}apek [Editor], ``\textit{The Concepts of Space and Time,
         Their Structure and Their Development}", Boston Studies in The Philosophy
         of Science, Vol. \textbf{74}, (Reidel Publishing Company, Boston, 1976).
\bibitem{Earman1987}J. Earman and J. Norton, ``What price spacetime
         substantivalism? The hole story'', \textit{British J. Phil.
         Sci.} \textbf{38} (1987), 515.
\bibitem{Berkley-1721}G. Berkley, ``De Motu or The principle and nature of motion and
         the cause of the communication of motions", (1721).
\bibitem{ref000}J.C. Maxwell, ``A dynamical theory of the electromagnetic field",
         \textit{Roy. Soc.} \textbf{155} (1865), 459. This article accompanied a December 8,
         1864 presentation by Maxwell to the Royal Society.
\bibitem{Adler-1966}I. Adler, ``\textit{A New Look at Geometry}", (John Day Company,
         NewYork, 1966).
\bibitem{Hardie-Aristotle}R.P. Hardie and R.K. Gaye [Translators], ``\textit{The Works of Aristotle}",
         Vol. \textbf{2} ``\textit{Physica}", (Clarendon Press, Oxford, 1930).
\bibitem{Aristotle}Aristotle, ``\textit{Physics}" (Oxford World's Classics), Edited
         by: D. Bostock, Translated by: R. Waterfield, (Oxford University Press, Oxford, 2008).
\bibitem{Mashhoon-1994}B. Mashhoon, H. Liu and P.S. Wesson, ``\textit{Space--time--matter}",
         In: ``\textit{Proceedings 7th Marcel Grossmann Meeting}", Stanford (1994), pp. 333--335.
\bibitem{d'Inverno-1992}R. d'Inverno, ``\textit{Introducing Einstein's Relativity}",
         (Clarendon Press, Oxford, 1992).
\bibitem{Janssen2005}M. Janssen, ``Of pots and holes: Einstein's bumpy road to general
         relativity'', \textit{Ann. Phys. (Berlin)} \textbf{14}
         Supplement, (2005), 58.
\bibitem{Mach1977}E. Mach, ``\textit{La Meccanica nel suo Sviluppo
         Storico--Critico (Mechanics in Its Development
         Historical--Critical)}'', (Boringhieri, Torino, 1977), Italian
         translation from the original 9th German Ed. of 1933 (1st
         Ed. 1883). Also published as ``{\it The Science of Mechanics: A Critical
         and Historical Account of Its Development}'', (Open Court,
         Illinois, 1960).
\bibitem{Barbour-1995}J. Barbour and H. Pfister [Editors], ``\textit{Mach's
         Principle: From Newton's Bucket to Quantum Gravity}'', Einstein
         Studies, Vol. \textbf{6}, (Birkh\"{a}user, Boston, 1995).
\bibitem{LenseThirring}J. Lense and H. Thirring, ``\"Uber den Einfluss der
         Eigenrotation der Zentralk\"orper auf die Bewegung der
         Planeten und Monde nach der Einsteinschen Gravitationstheorie (About the
         influence of the self--rotation of cenral body to the movement
         of planets and moons according to Einstein's theory of gravitation)'',
         {\it Physik. Z.}\ {\bf 19}\ (1918), 156.
\bibitem{Mashhoon1984}B. Mashhoon, F.W. Hehl and D.S. Theiss, ``On the gravitational
         effects of rotating masses: The Thirring--Lense papers'',
         \textit{Gen. Rel. Grav.}\ {\bf 16}\ (1984), 711.
\bibitem{Ciufolini2000}I. Ciufolini, ``The 1995--99 measurements of the Lense--Thirring
         effect using laser--ranged satellites'', \textit{Class. Quant. Grav.}\
         {\bf 17}\ (2000), 2369.
\bibitem{GP-B}F. Everitt, \textit{et al.}, ``Gravity Probe B: Final results of a space experiment
         to test general relativity'', \textit{Phys. Rev. Lett.}\
         {\bf 106}\ (2011), 221101.
\bibitem{Iorio2011}L. Iorio, ``Some considerations on the present--day results for the detection
         of frame--dragging after the final outcome of GP--B'',
         \textit{Europhys. Lett.}\ {\bf 96}\ (2011), 30001.
\bibitem{French-1966}A.P. French, ``\textit{Special Relativity}", (W.W. Norton, New
         York, 1966).
\bibitem{Resnick-1968}R. Resnick, ``\textit{Introduction to Special Relativity}",
         (Wiley, New York, 1968).
\bibitem{ref00-1}A.A. Michelson and E.W. Morley, ``On the relative motion of the
         earth and the luminiferous ether", \textit{Am. J. Sci.} \textbf{34} (1887), 333.
\bibitem{Shankland1}R.S. Shankland, S.W. McCuskey, F.C. Leone and G. Kuerti, ``New
         analysis of the interferometer observations of Dayton C.
         Miller", \textit{Rev. Mod. Phys.} \textbf{27} (1955), 167.
\bibitem{Shankland2}R.S. Shankland, ``Michelson--Morley experiment", \textit{Am. J. Phys.}
         \textbf{32} (1964), 16.
\bibitem{ref00-1-1}F. Fitzgerald, ``The ether and the earth's atmosphere",
         \textit{Sci.} \textbf{13} (1889), 390.
\bibitem{ref00-2}H.A. Lorentz, ``La th\'{e}orie \'{e}lectromagn\'{e}tique de Maxwell et son
         application aux corps mouvants (The electromagnetic theory of
         Maxwell and its application to moving bodies)", \textit{Arch. N\'{e}erl.
        Sci. Ex. Nat.} \textbf{25} (1892), 363.
\bibitem{ref00-3}H.A. Lorentz, ``The relative motion of the earth and the aether",
         \textit{Zitt. Akad. V. Wet.}  \textbf{1} (1892), 74.
\bibitem{Lorentz1909}H.A. Lorentz, ``{\it The Theory of Electrons and Its Applications
         to The Phenomena of Light and Radiatiant Heat}", (Columbia University Press,
         New York, 1909; Dover Publications, New York, 1952).
\bibitem{Lorentz1904}H.A. Lorentz, ``Electromagnetic phenomena in a system moving with
         any velocity less than that of light", \textit{Proc. Acad. Sci. Amsterdam}
         \textbf{6} (1904), 809. Reprinted in: ``{\it The Principle of
         Relativity: A Collection of Original Memoirs on The Special and General Theory of
         Relativity}'', by: H.A. Lorentz, A. Einstein, H. Minkowski and H. Weyl,
         Translated by: W. Perrett and G.B. Jeffery, (Dover Publications, New York,
         1952), pp. 9--34.
\bibitem{ref27-1}R.J. Kennedy and E.M. Thorndike, ``Experimental establishment of
         the relativity of time", \textit{Phys. Rev.} \textbf{42} (1932), 400.
\bibitem{ref0-1}J. Bradley, ``New discovered motion of the fixed stars",
         \textit{Phil. Trans. Roy. Soc.} \textbf{35} (1727), 637.
\bibitem{Stewart-1964}A.B. Stewart, ``The discovery of stellar aberration", \textit{Sci.
         Am.} \textbf{210} (March 1964), 100.
\bibitem{ref0-3}A.A. Michelson and E.W. Morley, ``Influence of motion of the
         medium on the velocity of light", \textit{Am. J. Sci.} \textbf{31}
         (1886), 377.
\bibitem{ref0-2}H.R. Bilger and W.K. Stowell, ``Light drag in a ring laser: An
         improved determination of the drag coefficient", \textit{Phys.
         Rev. A} \textbf{16} (1977), 313.
\bibitem{ref0-4}G.A. Sanders and S. Ezekiel, ``Measurement of Fresnel drag in
         moving media using a ring resonator technique", \textit{J. Opt.
         Soc. Am. B} \textbf{5} (1988), 674.
\bibitem{Ives}H.E. Ives, ``Historical note on the rate of a moving atomic clock",
         \textit{J. Opt. Soc. Am.} \textbf{37} (1947), 810.
\bibitem{Whittaker-1954}E.T. Whittaker, ``\textit{A History of The Theories of {\AE}ther
         and Electricity: The Modern Theories 1900--1926}",  (Nelson, London, 1953;
         Harper, New York, 1960; Humanities Press, London, 1973).
\bibitem{Holton}G. Holton, ``On the origins of the special theory of relativity",
         \textit{Am. J. Phys.} \textbf{28} (1960), 627.
\bibitem{Rindler70}W. Rindler, ``Einstein's priority in recognizing time dilation
         physically", \textit{Am. J. Phys.} \textbf{38} (1970), 1111.
\bibitem{Erlichson}H. Erlichson, ``The rod contraction--clock retardation ether theory and
         the special theory of relativity", \textit{Am. J. Phys.} \textbf{41} (1973), 1068.
\bibitem{Riess-1998}A.G. Riess, \textit{et al.}, ``Observational evidence from
         supernovae for an accelerating universe and a cosmological
         constant", \textit{Astron. J.} \textbf{116} (1998), 1009.
\bibitem{Perlmutter-1999}S. Perlmutter, \textit{et al.} [The Supernova Cosmology Project],
         ``Measurements of Omega and Lambda from $42$ high--redshift supernovae", \textit{Astrophys.
         J.} \textbf{517} (1999), 565.
\bibitem{supno3}A.G. Riess, {\it et al.}, ``BV RI light curves for $22$ type Ia supernovae",
        \textit{Astron. J.} \textbf{117} (1999), 707.
\bibitem{Riess2004}A.G. Riess, \textit{et al.}, ``Type Ia supernova discoveries at $ z>1$
        from the Hubble space telescope: Evidence for past deceleration and constraints
        on dark energy evolution", \textit{Astrophys. J.} \textbf{607} (2004), 665.
\bibitem{Einstein-1905}A. Einstein, ``Zur Elektrodynamik bewegter K\"{o}rper",
         \textit{Ann. Phys. (Berlin)} \textbf{322} (1905), 891. Its English version:
         ``On the electrodynamics of moving bodies", In: ``{\it The Principle of
         Relativity: A Collection of Original Memoirs on The Special and General Theory of
         Relativity}'', by: H.A. Lorentz, A. Einstein, H. Minkowski and H. Weyl,
         Translated by: W. Perrett and G.B. Jeffery, (Dover Publications, New York,
         1952), pp. 35--65.
\bibitem{Minkowski}H. Minkowski, ``Raum und Zeit'', \textit{Jber. Deutsch. Math.--Verein.}
        \textbf{18} (1909), 75. Address delivered at the 80th Assembly of German Natural
        Scientists and Physicians, Cologne, Sept. 21, 1908. Its English version: ``Space and
        time'', In: ``{\it The Principle of Relativity: A Collection of Original Memoirs on The
        Special and General Theory of Relativity}'', by: H.A. Lorentz, A. Einstein, H. Minkowski
        and H. Weyl, Translated by: W. Perrett and G.B. Jeffery, (Dover Publications, New York,
        1952), pp. 73--91.
\bibitem{Born}M. Born, ``Die Theorie des starren Elektrons in der Kinematik des
         Relativit\"{a}tsprinzips (The theory of rigid electron in the kinematics of
         principle of relativity)'', \textit{Ann. Phys. (Berlin)} \textbf{335} (1909), 1.
\bibitem{Ehrenfest}P. Ehrenfest, ``Gleichf\"{o}rmige Rotation starrer K\"{o}rper
         und Relativit\"{a}tstheorie (Uniform rotation of rigid bodies and theory of
         relativity.)'', \textit{Physik. Z.} \textbf{10} (1909), 918.
\bibitem{Petkov2005}V. Petkov, ``\textit{Relativity and The Nature of
         Spacetime}'', (Springer, Berlin, 2005).
\bibitem{Iorio2006}A. Iorio, ``Three questions on Lorentz violation'',
         \textit{J. Phys. Conf. Ser.} \textbf{67} (2007), 012008.
\bibitem{Mattingly-2005}D. Mattingly, ``Modern tests of Lorentz invariance",
         \textit{Living Rev. Rel.} \textbf{8} (2005), 5.
\bibitem{ref5}K. Shinozaki, {\it et al.} [AGASA Collaboration], ``AGASA
         results", \textit{Nucl. Phys. B} \textbf{136} (2004), 18.
\bibitem{ref5-1}F.W. Stecker, M.A. Malkan and S.T. Scully, ``Intergalactic photon
         spectra from the far--IR to the UV Lyman limit for $0<z<6$ and
         the optical depth of the universe to high--energy gamma rays",
         \textit{Astrophys. J.} \textbf{648} (2006), 774.
\bibitem{ref6}R.U. Abbasi, {\it et al.}, ``First observation of the
         Greisen--Zatsepin--Kuzmin suppression", \textit{Phys. Rev. Lett.}
         \textbf{100} (2008), 101101.
\bibitem{Drozdzynski2011}J. Dro\.{z}d\.{z}y\'{n}ski, ``Evidence for an invalidity
         of the principle of relativity'', \textit{J. Mod. Phys.} \textbf{2} (2011), 1247.
\bibitem{Sela2009}O. Sela, B. Tamir, S. Dolev and A.C. Elitzur, ``Can special relativity
         be derived from Galilean mechanics alone?'', \textit{Found. Phys.}
         \textbf{39} (2009), 499.
\bibitem{ColemanGlashow}S. Coleman and S.L. Glashow, ``High--energy tests of Lorentz
         invariance", \textit{Phys. Rev. D} \textbf{59} (1999), 116008.
\bibitem{Einstein-1915}A. Einstein, ``Die Feldgleichungen der Gravitation (The field
         equations of gravitation)", \textit{Preuss. Akad. Wiss. Berlin
         Sitz.} \textbf{17} (1915), 844.
\bibitem{Einstein-1916}A. Einstein, ``Die Grundlage der allgemeinen
         Relativit\"{a}tstheorie", \textit{Ann. Phys. (Berlin)} \textbf{354} (1916), 769.
         Its English version: ``The foundation of the general theory of
         relativity", In: ``{\it The Principle of Relativity: A Collection
         of Original Memoirs on The Special and General Theory of
         Relativity}'', by: H.A. Lorentz, A. Einstein, H. Minkowski and H. Weyl,
         Translated by: W. Perrett and G.B. Jeffery, (Dover Publications, New York,
         1952), pp. 109--164.
\bibitem{Lichtenegger-2005}H. Lichtenegger and B. Mashhoon, ``Mach's principle", In:
         ``\textit{The Measurment of Gravitomagnetism: A Challenging
         Enterprise}", Edited by: L. Iorio, (NOVA Science,
         Hauppage, New York, 2005), pp. 13--27, \textit{arXiv: physics/0407078}.
\bibitem{CiufoliniWheeler1995}I. Ciufolini and J.A. Wheeler, ``{\it Gravitation and
         Inertia}'', (Princeton University Press, Princeton, 1995).
\bibitem{Straumann2004}N. Straumann, ``{\it General Relativity
         With Applications to Astrophysics}'', (Springer, Berlin,
         2004).
\bibitem{Mehra}J. Mehra, ``{\it Einstein, Hilbert, and The Theory of Gravitation}'',
         (Reidel Publishing Company, Holland, 1974).
\bibitem{Pais}A. Pais, ``{\it Subtle Is The Lord, The Science and The Life of
         Albert Einstein}'', (Oxford University Press, Oxford, 1982).
\bibitem{Stachel-1912}J. Stachel, ``Einstein's struggle with general covariance,
         1912--1915", Presented at General Relativity and Gravitation 9th,
         1980 at Jena, Germany; Reprinted as ``Einstein's search for
         general covariance, 1912--1915", In: ``\textit{Einstein
         and The History of General Relativity}", based on the
         Proceedings of May 1986, Osgood Hill Conference, Massachusetts,
         Edited by: D. Howard and J. Stachel, (The Center for Einstein
         Studies, Boston University, 1989), pp. 63--100.
\bibitem{Stachel}J. Stachel, ``What a physicist can learn from the discovery of
         general relativity'', In: ``{\it Proceedings of The Fourth Marcel
         Grossmann Meeting on General Relativity}'', Edited by: R.
         Ruffini, (North--Holland, Amsterdam, 1986), pp. 1857--1862.
\bibitem{Norton}J. Norton, ``How Einstein found his field equations,
         1912--1915'', In: ``{\it Einstein and The History of General Relativity}'',
         based on the Proceedings of May 1986, Osgood Hill Conference,
         Massachusetts,
         Edited by: D. Howard and J. Stachel, (The Center for Einstein
         Studies, Boston University, 1989), pp. 101--159. It is reprinted
         from ``{\it Historical Studies in The Physical Sciences}'', Vol. \textbf{14},
         Part~2, Edited by: J.L. Heilbron, (The Regents of The University of
         California, Berkeley, 1984), pp. 253--316.
\bibitem{Synge1960}J.L. Synge, ``{\it Relativity: The General Theory}'', (North--Holland,
         Amsterdam, 1960).
\bibitem{bida}N.D. Birrell and P.C.W. Davies, ``{\it Quantum Fields in Curved
         Space}'', (Cambridge University Press, Cambridge, 1982).
\bibitem{bos}I.L. Buchbinder, S.D. Odintsov and I.L. Shapiro, ``{\it
         Effective Action in Quantum Gravity}'', (Institute of Physics
         Publishing, Bristol, 1992).
\bibitem{Einstein-1913}A. Einstein and M. Grossmann, ``Entwurf einer verallgemeinerten
         Relativit\"{a}tstheorie und einer Theorie der Gravitation (Draft
         of a generalized relativity theory and a theory of gravitation)",
         \textit{Z. Math. Phys.} \textbf{62} (1913), 225.
\bibitem{Einstein-1914}A. Einstein and M. Grossmann, ``Kovarianzeigenschaften der
         Feldgleichungen der auf die verallgemeinerte Relativit\"{a}tstheorie gegr\"{u}ndeten
         Gravitationstheorie (Covariance properties of the field equations of the
         gravitational theory based on generalized relativity)", \textit{Z. Math. Phys.}
         \textbf{63} (1914), 215.
\bibitem{Einstein-Ehrenfest}A. Einstein wrote to: P. Ehrenfest, on 26th
         December, 1915, EA 9--363.
\bibitem{Einstein-Besso}A. Einstein wrote to: M. Besso, on 3rd January, 1916, In:
         ``\textit{Albert Einstein, Michele Besso Correspondence
         1903--1955}'', Edited by: P. Speziali, (Hermann, Paris, 1972), pp. 63--64.
\bibitem{Einstein-1952}A. Einstein, ``\textit{Relativity and The Problem of Space
         (1952)}", Appendix~$5$, In: ``\textit{Relativity, The Special and
         The General Theory: A Popular Exposition}", Translated by: R.W.
         Lawson, (Methuen, London, 15th Ed. 1954), pp. 135--157.
\bibitem{Eddington-1921}A.S. Eddington, `` `Space' or `{\AE}ther'?", \textit{Nature}
         \textbf{107} (1921), 201.
\bibitem{ref27-4}A. Einstein, ``\textit{\"{A}ther und Relativit\"{a}tstheorie (Ether
         and Relativity Theory)}", (Springer, Berlin, 1920), reprinted as
         ``\textit{Sidelights on Relativity}", (Dover Publications, New York, 1983).
\bibitem{Trautman-1966}A. Trautman,``Comparison of Newtonian and relativistic theories of
         space--time", In: ``\textit{Perspectives in Geometry and
         Relativity}", Edited by: B. Hoffmann, (Indiana University Press,
         Bloomington, 1966), pp. 413--425.
\bibitem{Whittaker-1951}E.T. Whittaker, ``\textit{A History of The Theories of {\AE}ther and
         Electricity: The Classical Theories}", (Nelson, London, 2nd Ed. 1951;
         Tomash Publishers, New York, 1987).
\bibitem{Schaffner-1972}K.F. Schaffner, ``\textit{Nineteenth--Century {\AE}ther Theories}",
         (Pergamon Press, New York, 1972).
\bibitem{Dupre-2012}M.J. Dupr\'{e} and F.J. Tipler, ``General relativity as an {\ae}ther
         theory", \textit{Int. J. Mod. Phys. D} \textbf{21} (2012), 1250011.
\bibitem{Gautreau-2000}R. Gautreau, ``Newton's absolute time and space in general
         relativity", \textit{Am. J. Phys.} \textbf{68} (2000), 350.
\bibitem{Savickas}D. Savickas, ``General relativity exactly described in terms of Newton's
         laws within curved geometries'', \textit{Int. J. Mod. Phys. D}
         \textbf{23} (2014), 1430018.
\bibitem{Salam-1980}A. Salam, ``Gauge unification of fundamental forces",
         \textit{Rev. Mod. Phys.} \textbf{52} (1980), 525.
\bibitem{Wesson-1992}P.S. Wesson and J. Ponce de Leon, ``Kaluza--Klein equations,
         Einstein's equations, and an effective energy--momentum
         tensor", \textit{J. Math. Phys.} \textbf{33} (1992), 3883.
\bibitem{Romero-1996}C. Romero, R. Tavakol and R. Zalaletdinov, ``The embedding of
         general relativity in five dimensions", \textit{Gen. Rel. Grav.}
         \textbf{28} (1996), 365.
\bibitem{Overduin-1997}J.M. Overduin and P.S. Wesson, ``Kaluza--Klein gravity",
         \textit{Phys. Rep.} \textbf{283} (1997), 303.
\bibitem{Wesson-1999}P.S. Wesson, ``\textit{Space--Time--Matter: Modern Kaluza--Klein
         Theory}", (World Scientific, Singapore, 1999).
\bibitem{Wesson-2006}P.S. Wesson, ``\textit{Five--Dimensional Physics:
         Classical and Quantum Consequences of Kaluza--Klein Cosmology}", (World
         Scientific, Singapore, 2006).
\bibitem{Bahrehbakhsh-2010}A.F. Bahrehbakhsh, M. Farhoudi and H. Shojaie, ``FRW cosmology
         from five dimensional vacuum Brans--Dicke theory", \textit{Gen.
         Rel. Grav.} \textbf{43} (2010), 847.
\bibitem{Rasouli-2011}S.M.M. Rasouli, M. Farhoudi and H.R. Sepangi ``Anisotropic
         cosmological model in modified Brans--Dicke theory",
         \textit{Class. Quant. Grav.} \textbf{28} (2011), 155004.
\bibitem{Bahrehbakhsh-2013}A.F. Bahrehbakhsh, M. Farhoudi and H. Vakili, ``Dark energy from
         fifth dimensional Brans--Dicke theory", \textit{Int. J. Mod.
         Phys. D} \textbf{22} (2013), 1350070.
\bibitem{Rasouli-2014}S.M.M. Rasouli, M. Farhoudi and P.V. Moniz, ``Modified
         Brans--Dicke  theory in arbitrary dimensions", \textit{Class.
         Quant. Grav.} \textbf{31} (2014), 115002.
\bibitem{Harko-2008}T. Harko, ``Modified gravity with arbitrary coupling between
         matter and geometry", \textit{Phys. Lett. B} \textbf{669}
         (2008), 376.
\bibitem{Harko-2011}T. Harko, F.S.N. Lobo, S. Nojiri and S.D. Odintsov, ``$f(R,T)$
         gravity", \textit{Phys. Rev. D} \textbf{84} (2011), 024020.
\bibitem{Bisabr-2012}Y. Bisabr, ``Modified gravity with a nonminimal gravitational
         coupling to matter", \textit{Phys. Rev. D} \textbf{86} (2012), 044025.
\bibitem{Jamil2012}M. Jamil, D. Momeni, R. Muhammad and M. Ratbay, ``Reconstruction of some
        cosmological models in $f(R,T)$ gravity", \textit{Eur. Phys. J. C}
        \textbf{72} (2012), 1999.
\bibitem{Alvarenga2013}F.G. Alvarenga, A. de la Cruz--Dombriz, M.J.S. Houndjo, M.E. Rodrigues and
        D. Saez--Gomez, ``Dynamics of scalar perturbations in $f(R,T)$ gravity",
        \textit{Phys. Rev. D}  \textbf{87} (2013), 103526.
\bibitem{Haghani-2013}Z. Haghani, T. Harko, F.S.N. Lobo,  H.R. Sepangi and S. Shahidi,
         ``Further matters in space--time geometry: $f(R,T,R_{\mu\nu}T^{\mu\nu})$
         gravity", \textit{Phys. Rev. D} \textbf{88} (2013), 044023.
\bibitem{Shabani-2013}H. Shabani and M. Farhoudi, ``$f(R,T)$ cosmological models in
         phase--space", \textit{Phys. Rev. D} \textbf{88} (2013), 044048.
\bibitem{Shabani-0000}H. Shabani and M. Farhoudi, ``Cosmological and solar system
         consequences of $f(R,T)$ gravity models'', \textit{Phys. Rev. D}
         \textbf{90} (2014), 044031.
\bibitem{ZareFarhoudi}R. Zaregonbadi and M. Farhoudi, ``Late time acceleration from
         matter--curvature coupling'', submitted to journal.
\bibitem{EinsteinCC}A. Einstein, ``Kosmologische Betrachtungen zur allgemeinen
         Relativit\"atstheorie'', {\it Preuss. Akad. Wiss. Berlin, Sitz.} (1917),
         142. Its English version: ``Cosmological considerations on the general
         theory of relativity'', In: ``{\it The Principle of Relativity: A Collection
         of Original Memoirs on The Special and General Theory of
         Relativity}'', by: H.A. Lorentz, A. Einstein, H. Minkowski and H. Weyl,
         Translated by: W. Perrett and G.B. Jeffery, (Dover Publications, New York,
         1952), pp. 175--188.
\bibitem{deSitter}W. de Sitter, ``On the curvature of space", \textit{Proc. Kon. Ned.
         Acad. Wet.} \textbf{20} (1918), 229.
\bibitem{Hubble1929}E.P. Hubble, ``A relation between distance and radial
         velocity among extragalactic nebulae'', \textit{Proc. Nat. Acad.
         Sci. USA} \textbf{15} (1929), 169.
\bibitem{Gamov1970}G. Gamow, ``\textit{My World Line, An Informal
         Autobiography}'', (Viking, New York, 1970).
\bibitem{Pirani}A. Einstein wrote to: F. Pirani, 1954, EA 17--448.
\bibitem{NamFar}N. Namavarian and M. Farhoudi, ``Cosmological constant implementing
         Mach principle in general relativity", submitted to journal.
\bibitem{Cos.pro1}S. Weinberg, ``The cosmological constant problem'', \textit{Rev. Mod.
         Phys.} \textbf{61} (1989), 1.
\bibitem{Carroll}S.M. Carroll, ``The cosmological constant", \textit{Living. Rev. Rel.}
         \textbf{4} (2001), 1.
\bibitem{Sahni}V. Sahni, ``The cosmological constant problem and quintessence",
         \textit{Class. Quant. Grav.} \textbf{19} (2002), 3435.
\bibitem{Nobbenhuis}S. Nobbenhuis, ``Categorizing different approaches to the cosmological
         constant problem'', \textit{Found. Phys.} \textbf{36} (2006), 613.
\bibitem{Padmanabhan}H. Padmanabhan and T. Padmanabhan, ``CosMIn: The solution to the cosmological
         constant problem", \textit{Int. J. Mod. Phys. D} \textbf{22} (2013), 1342001.
\bibitem{Bernard}D. Bernard and A. LeClair, ``Scrutinizing the cosmological constant problem
         and a possible resolution", \textit{Phys. Rev. D} \textbf{87} (2013), 063010.
\bibitem{WeinbergBook}S. Weinberg, ``\textit{Cosmology}'', (Oxford University Press, Oxford, 2008).
\bibitem{Barrow2011}J.D. Barrow and D.J. Shaw, ``The value of the cosmological constant",
         \textit{Gen. Rel. Grav.} \textbf{43} (2011), 2555.
\bibitem{Peebles-2003}P.J.E. Peebles and B. Ratra, ``The cosmological constant and dark energy",
         \textit{Rev. Mod. Phys.} \textbf{75} (2003), 559.
\bibitem{Padmanabhan2003}T. Padmanabhan, "Cosmological constant--the weight of the vacuum",
         \textit{Phys. Rep.} \textbf{380} (2003), 235.
\bibitem{Polarski-2006}D. Polarski, ``Dark energy: Current issues", \textit{Ann.
         Phys. (Berlin)} \textbf{15} (2006), 342.
\bibitem{Copeland-2006}E.J. Copeland, M. Sami and S. Tsujikawa, ``Dynamics of dark
         energy", \textit{Int. J. Mod. Phys. D} \textbf{15} (2006), 1753.
\bibitem{Durrer-2008}R. Durrer and R. Maartens, ``Dark energy and dark gravity: Theory
         overview", \textit{Gen. Rel. Grav.} \textbf{40} (2008), 301.
\bibitem{Bamba-2012}K. Bamba, S. Capozziello, S. Nojiri and S.D. Odintsov, ``Dark
         energy cosmology: The equivalent description via different
         theoretical models and cosmography tests", \textit{Astrophys.
         Space Sci.} \textbf{342} (2012), 155.
\bibitem{Ade-2013}P.A.R. Ade, {\it et al.} [Planck Collaboration], ``Planck 2013 results. XVI.
         Cosmological parameters", \textit{Astron. Astrophys.} \textbf{571} (2014), A16.
\bibitem{Ade-2015}P.A.R. Ade, {\it et al.} [Planck Collaboration], ``Planck 2015 results. XIII.
         Cosmological parameters", \textit{arXiv: 1502.01589}.
\bibitem{Farhoudi-2006}M. Farhoudi, ``On higher order gravities, their analogy to GR, and
         dimensional dependent version of Duff's trace anomaly
         relation", \textit{Gen. Rel. Grav.} \textbf{38} (2006), 1261.
\bibitem{dmatt1}G. Bertonea, D. Hooperb and J. Silk, ``Particle dark matter: Evidence,
         candidates and constraints", \textit{Phys. Rep.} \textbf{405} (2005), 279.
\bibitem{dmatt2}J. Silk, ``Dark matter and galaxy formation", \textit{Ann. Phys.
         (Berlin)} \textbf{15} (2006), 75.
\bibitem{dmatt3}J.L. Feng, ``Dark matter candidates from particle physics and methods
         of detection", \textit{Annu. Rev. Astron. Astrophys.} \textbf{48} (2010), 495.
\bibitem{dmatt4}L. Bergstr\"{o}m, ``Dark matter evidence, particle physics candidates
         and detection methods", \textit{Ann. Phys. (Berlin)} \textbf{524} (2012), 479.
\bibitem{dmatt5}C.S. Frenk and S.D.M. White, ``Dark matter and cosmic structure",
         \textit{Ann. Phys. (Berlin)} \textbf{524} (2012), 507.
\bibitem{farc}M. Farhoudi, ``Classical trace anomaly'', {\it Int. J. Mod.
         Phys. D}\ {\bf 14}\ (2005), 1233.
\bibitem{Brans-1961}C. Brans and R.H. Dicke, ``Mach's principle and a relativistic
         theory of gravitation", \textit{Phys. Rev.} \textbf{124} (1961), 925.
\bibitem{Dicke-1962}R.H. Dicke, ``Mach's principle and invariance under
         transformation of units", \textit{Phys. Rev.} \textbf{125} (1962), 2163.
\bibitem{Fujii-2004}Y.  Fujii and  K. Maeda,  ``\textit{The Scalar--Tensor Theory of
         Gravitation}", (Cambridge University Press, Cambridge 2004).
\bibitem{Farajollahi-2010}H. Farajollahi, M. Farhoudi and H. Shojaie, ``On dynamics of
         Brans--Dicke theory of gravitation", \textit{Int. J. Theor. Phys.}
         \textbf{49}  (2010), 2558.
\bibitem{Faraoni-2004}V. Faraoni, ``\textit{Cosmology in Scalar Tensor Gravity}",
         (Kluiwer Academic Publishers, Netherlands, 2004).
\bibitem{Capozziello-2011}S. Capozziello and V. Faraoni, ``\textit{Beyond Einstein Gravity:
         A Survey of Gravitational Theories for Cosmology and
         Astrophysics}", (Springer, Heidelberg, 2011).
\bibitem{Khoury-2004}J. Khoury and A. Weltman, ``Chameleon cosmology", \textit{Phys.
         Rev. D} \textbf{69}  (2004), 044026.
\bibitem{Brax-2010}P. Brax, C. Burrage, A.--C. Davis, D. Seery and A. Weltman,
         ``Higgs production as a probe of chameleon dark energy",
         \textit{Phys. Rev. D} \textbf{81} (2010), 103524.
\bibitem{Farajollahi-2012}H. Farajollahi, M. Farhoudi, A. Salehi and H. Shojaie,
         ``Chameleonic generalized Brans--Dicke model and late--time
         acceleration", \textit{Astrophys. Space Sci.} \textbf{337} (2012), 415.
\bibitem{SabaFarhoudi}N. Saba and M. Farhoudi, ``Chameleonic inflation in the light of
         Planck 2015'', work in progress.
\bibitem{JacMat2001}T. Jacobson and D. Mattingly, ``Gravity with a dynamical preferred
         frame", \textit{Phys. Rev. D} \textbf{64} (2001), 024028.
\bibitem{Eling-2004}C. Eling, T. Jacobson and D. Mattingly, ``Einstein--{\ae}ther
         theory", \textit{arXiv: gr--qc/0410001}.
\bibitem{Jacobson-2007}T. Jacobson, ``Einstein--{\ae}ther gravity: A status report",
         \textit{PoS QG--Ph} (2007), 020, \textit{arXiv: 0801.1547}.
\bibitem{barow2012}J.D. Barrow, ``Some inflationary Einstein--aether cosmologies'',
         \textit{Phys. Rev. D} \textbf{85} (2012), 047503.
\bibitem{Wei2014}H. Wei, X.--P. Yan and Y.--N. Zhou, ``Cosmological evolution of
         Einstein--aether models with power--law--like potential'',
         \textit{Gen. Rel. Grav.} \textbf{46} (2014), 1719.
\bibitem{Haghani2014}Z. Haghani, T. Harko, H.R. Sepangi and S. Shahidi, ``Scalar
         Einstein-aether theory'', \textit{arXiv: 1404.7689}.
\bibitem{Furtado-2013}C. Furtado, J.R. Nascimento, A.Y. Petrov and A.F. Santos, ``The
         {\ae}ther--modified gravity and the G\"{o}del metric", \textit{arXiv: 1109.5654}.
\bibitem{Archer-Plato}``\textit{The Timaeus of Plato}", Edited with Introduction and
         Notes by: R.D. Archer--Hind, (Macmillan, London, 1888).
\bibitem{Plato}Plato, ``\textit{Timaeus}", Translated by: B. Jowett, (Echo
         Library, United Kingdom, 2006).
\bibitem{Webb-1999}J.K. Webb, {\it et al.}, ``A search for time variation of the fine
         structure constant", \textit{Phys. Rev. Lett.} \textbf{82} (1999), 884.
\bibitem{ref1}M.T. Murphy, {\it et al.}, ``Possible evidence for a variable fine
         structure constant from QSO absorption lines: Motivations,
         analysis and results", \textit{Mon. Not. Roy. Astron. Soc.}
         \textbf{327} (2001), 1208.
\bibitem{Albert}J. Albert, \textit{et al.} [MAGIC Collaboration], ``Probing quantum gravity
         using photons from a flare of the active galactic nucleus Markarian 501 observed
         by the MAGIC telescope", \textit{Phys. Lett. B} \textbf{668} (2008), 253.
\bibitem{Barrow-1998}J.D. Barrow and J. Magueijo, ``Varying--$\alpha$ theories and
         solutions to the cosmological problems", \textit{Phys. Lett. B}
         \textbf{443} (1998), 104.
\bibitem{ref2}H.B. Sandvik, J.D. Barrow and J. Magueijo, ``A simple
         varying--alpha cosmology", \textit{Phys. Rev. Lett.} \textbf{88}
         (2002), 031302.
\bibitem{Barrow-Magueijo-1998}J.D. Barrow and J. Magueijo, ``Solutions to the
         quasi--flatness and quasi--lambda problems'', \textit{Phys. Lett. B}
         {\bf447} (1998), 246.
\bibitem{Clayton-Moffat-1998}M.A. Clayton and J.W. Moffat, ``Dynamical
         mechanism for varying light velocity as a solution to cosmological
         problem'', \textit{Phys. Lett. B} {\bf 480} (1998), 263.
\bibitem{Albrecht-Magueijo-1999}A. Albrecht and J. Magueijo, ``A time varying
         speed of light as a solution to cosmological puzzles'', \textit{Phys.
         Rev. D} {\bf 59} (1999), 043516.
\bibitem{Magueijo}J. Magueijo, ``New varying speed of light theories'', \textit{Rep.
         Prog. Phys.} {\bf 66} (2003), 2025.
\bibitem{ref4-3}H. Shojaie and M. Farhoudi, ``A cosmology with variable c",
         \textit{Can. J. Phys.} \textbf{84} (2006), 933.
\bibitem{ref4-5}H. Shojaie and M. Farhoudi, ``A varying--c cosmology",
         \textit{Can. J. Phys.} \textbf{85} (2007), 1395.
\bibitem{ref3}J. Magueijo, J.D. Barrow and H.B. Sandvik, ``Is it e or is it c?
         Experimental tests of varying alpha", \textit{Phys. Lett. B}
         \textbf{549} (2002), 284.
\bibitem{ref18}P. Castorina and D. Zappala, ``Noncommutative electrodynamics and
         ultra high energy gamma rays", \textit{Europhys. Lett.}
         \textbf{64} (2003), 641.
\bibitem{ref19}R. Horvat, D. Kekez, P. Schupp, J. Trampeti and J. You,
         ``Photon--neutrino interaction in  $\theta$--exact covariant
         noncommutative field theory", \textit{Phys. Rev. D} \textbf{84}
         (2011), 045004.
\bibitem{ref10}G. Amelino--Camelia, ``Relativity in space--times with
         short--distance structure governed by an observer--independent
         (Planckian) length scale", \textit{Int. J. Mod. Phys. D}
         \textbf{11} (2002), 35.
\bibitem{ref12}J. Magueijo and L. Smolin, ``Lorentz invariance with an invariant
         energy scale", \textit{Phys. Rev. Lett.} \textbf{88} (2002), 190403.
\bibitem{ref11}J. Kowalski--Glikman and S. Nowak, ``Non--commutative space--time
         of doubly special relativity theories", \textit{Int. J. Mod. Phys.
         D} \textbf{12} (2003), 299.
\bibitem{ref13}H.--Y. Guo, C.--G. Huang, Z. Xu and B. Zhou, ``On de Sitter
         invariant special relativity and cosmological constant as origin
         of inertia", \textit{Mod. Phys. Lett. A} \textbf{19} (2004), 1701.
\bibitem{ref16}A. Agostini, G. Amelino--Camelia and F. D'Andrea, ``Hopf--algebra
         description of noncommutative--spacetime symmetries", \textit{Int.
         J. Mod. Phys. A} \textbf{19} (2004), 5187.
\bibitem{ref14}H.--Y. Guo, H.--T. Wu and B. Zhou, ``The principle of relativity
         and the special relativity triple", \textit{Phys. Lett. B}
         \textbf{670} (2009), 437.
\bibitem{ref17}G. Amelino--Camelia, ``Doubly--special relativity: Facts, myths
         and some key open issues", \textit{Symmetry} \textbf{2} (2010), 230.
\bibitem{Alexander-2000}S. Alexander, ``On the varying speed of light in a brane--induced
         FRW universe", \textit{J. High Energy Phys.} \textbf{0011} (2000), 017.
\bibitem{ref23}I.T. Drummond and S.J. Hathrell, ``QED vacuum polarization in a
         background gravitational field and its effect on the velocity of
         photons", \textit{Phys. Rev. D} \textbf{22} (1980), 343.
\bibitem{ref26}M.A. Clayton and J.W. Moffat, ``Dynamical mechanism for varying
         light velocity as a solution to cosmological problems",
         \textit{Phys. Lett. B} \textbf{460} (1999), 263.
\bibitem{ref25}M.A. Clayton and J.W. Moffat, ``Scalar--tensor gravity theory for
         dynamical light velocity", \textit{Phys. Lett. B} \textbf{477}
         (2000), 269.
\bibitem{ref24}J. Magueijo, ``Bimetric varying speed of light theories and
         primordial fluctuations", \textit{Phys. Rev. D} \textbf{79}
         (2009), 043525.
\bibitem{ref7-1}D. Finkbeiner, M. Davis and D. Schlegel, ``Detection of a far IR
         excess with DIRBE at 60 and 100 microns", \textit{Astrophys. J.}
         \textbf{544} (2000), 81.
\bibitem{ref7}D. Mazin and M. Raue, ``New limits on the density of the
         extragalactic background light in the optical to the far infrared
         from the spectra of all known TeV blazars", \textit{Astron.
         Astrophys.} \textbf{471} (2007), 439.
\bibitem{ref7-2}A.A. Penzias and R.H. Wilson, ``A measurement of excess antenna
         temperature at 4080 Mc/s", \textit{Astrophys. J.} \textbf{142}
         (1965), 419.
\bibitem{ref8}K. Greisen, ``End to the cosmic--ray spectrum", \textit{Phys. Rev.
         Lett.} \textbf{16} (1966), 748.
\bibitem{ref9}G.T. Zatsepin and V.A. Kuzmin, ``Upper limit of the spectrum of
         cosmic rays", \textit{J. Exp. Theor. Phys. Lett.} \textbf{4}
         (1966), 78.
\bibitem{Majid2000}S. Majid, ``Foundations of Quantum Group Theory'', (Cambridge
         University Press, Cambridge, 2000).
\bibitem{Fock1964}V.A. Fock, ``The Theory of Space--Time and Gravitation'', (Pergamon
         Press, New York, 1964).
\bibitem{Hossenfelder}S. Hossenfelder, ``The box--problem in deformed special
         relativity'', {\it arXiv: 0912.0090}.
\bibitem{ref41}S.T. Scully and F.W. Stecker, ``Lorentz invariance violation and
         the observed spectrum of ultrahigh energy cosmic rays",
         \textit{Astropart. Phys.} \textbf{31} (2009), 220.
\bibitem{YouFar}M. Yousefian and M. Farhoudi, ``Justification of Webb's redshift and ultra
         high energy cosmic rays via an ether model", work in progress.
\end{thebibliography}
\end{document}